\documentclass[11pt]{article}
\usepackage{amsmath,amssymb,amsfonts,latexsym,amsthm,enumerate,fullpage}
\newtheorem{thm}{Theorem}
\newtheorem*{thm*}{Theorem}
\newtheorem{lemma}[thm]{Lemma}
\newtheorem*{lemma*}{Lemma}

\newtheorem*{prop*}{Proposition}
\newtheorem{cor}[thm]{Corollary}

\newtheorem{claim}[thm]{Claim}
\theoremstyle{remark}

\newtheorem*{rmk*}{Remark}
\newtheorem*{rmks*}{Remarks}
\newtheorem*{not*}{Notation}
\newtheorem*{claim*}{Claim}
\newtheorem*{fact*}{Fact}
\newtheorem*{conj*}{Conjecture}

\newtheorem*{dfn*}{Definition}

\def\R{\mathbb{R}}
\def\E{\mathbb{E}}
\def\P{\mathbb{P}}

\def\PP{\mathcal{P}}
\def\Cvar{\mathcal{C}^{\textrm{var}}}
\def\Crow{\mathcal{C}^{\textrm{disc}}}

\def\Var{\mathrm{Var}}
\def\N{\mathcal{N}}
\def\O{\mathcal{V}}

\newcommand\ignore[1]{}
\newcommand\ip[1]{\left<#1\right>}

\newcommand{\calS}{\mathcal{S}}
\newcommand{\dpm}{\{1,-1\}}
\newcommand{\rgta}{\rightarrow}

\newif\ifnotes\notestrue

\ifnotes

\newcommand{\rnote}[1]{{\bf (Raghu:} {#1}{\bf ) }}
\newcommand{\snote}[1]{{\bf (Shachar:} {#1}{\bf ) }}
\newcommand{\onote}[1]{{\bf (Oded:} {#1}{\bf ) }}

\else

\newcommand{\rnote}[1]{}
\newcommand{\snote}[1]{}
\newcommand{\onote}[1]{}

\fi

\date{}
\begin{document}

\title{Constructive Discrepancy Minimization by Walking on The Edges}
\author{Shachar Lovett\thanks{Supported by NSF grant DMS-0835373.}\\
Institute for Advanced Study\\
\texttt{slovett@math.ias.edu}
\and
Raghu Meka\thanks{Supported by NSF grants DMS-0835373 and CCF-0832797.}\\
Institute for Advanced Study\\
\texttt{raghu@math.ias.edu}
}
\maketitle{}

\begin{abstract}
Minimizing the discrepancy of a set system is a fundamental problem in combinatorics. One of the cornerstones in this area is the celebrated six standard deviations result of Spencer (AMS 1985): In any system of $n$ sets in a universe of size $n$, there always exists a coloring which achieves discrepancy $6\sqrt{n}$. The original proof of Spencer was existential in nature, and did not give an efficient algorithm to find such a coloring. Recently, a breakthrough work of Bansal (FOCS 2010) gave an efficient algorithm which finds such a coloring. His algorithm was based on an SDP relaxation of the discrepancy problem and a clever rounding procedure. In this work we give a new randomized algorithm to find a coloring as in Spencer's result based on a restricted random walk we call {\sl Edge-Walk}. Our algorithm and its analysis use only basic linear algebra and is ``truly'' constructive in that it does not appeal to the existential arguments, giving a new proof of Spencer's theorem and the {\sl partial coloring lemma}.
\end{abstract}
\newcommand{\disc}{\mathsf{disc}}
\section{Introduction}
Minimizing the discrepancy of a set system is a fundamental problem in combinatorics with many applications in computer science (see \cite{Matousek1999, Chazelle2002}). Here, we are given a collection of sets $\calS$ from a universe $V = \{1,\ldots,n\}$ and the goal is to find a {\it coloring} $\chi:V \rightarrow \dpm$ that minimizes the maximum discrepancy $\chi(\calS) = \max_{S \in \calS} |\sum_{i \in S} \chi(i)|$. We denote the minimum discrepancy of $\calS$ by $\disc(\calS)$.

There is by now a rich body of literature on discrepancy minimization with special focus on the `discrete' formulation described above. One of the cornerstones in this area is the celebrated six standard deviations result of Spencer~\cite{Spencer85}.
\begin{thm}
  For any set system $(V,\calS)$ with $|V| = n$, $|\calS| = m$, there exists a coloring $\chi:V \rgta \dpm$ such that $\chi(\calS) < K \sqrt{n \cdot \log_2(m/n)}$, where $K$ is a universal constant ($K$ can be $6$ if $m=n$).
\end{thm}

One remarkable aspect of the above theorem is that for $m = O(n)$, the discrepancy is just $O(\sqrt{n})$, whereas a random coloring has discrepancy $O(\sqrt{n \log n})$. Spencer's original proof relied on an ingenious pigeon-hole principle argument based on Beck's partial coloring approach \cite{Beck81}. However, due to the use of the pigeon-hole principle, the proof was non-constructive: Spencer's proof does not give an efficient (short of enumerating all possible colorings) way to find a good coloring $\chi$ as in the theorem. This was a longstanding open problem in discrepancy minimization and it was even conjectured that such an algorithm cannot exist \cite{AlonS11}. In a recent breakthrough work, Bansal~\cite{Bansal10} disproved this conjecture and gave the first randomized polynomial time algorithm to find a coloring with discrepancy $O(\sqrt{n} \cdot \log(m/n))$, thus matching Spencer's bound up to constant factors for the important case of $m = O(n)$.

In this work we give a new elementary constructive proof of Spencer's result. Our algorithm and its analysis use only basic linear algebra and perhaps more importantly is ``truly'' constructive. Bansal's algorithm while giving a constructive solution, still implicitly uses Spencer's original non-constructive proof to argue the correctness of the algorithm. Our algorithm on the other hand also gives a new (constructive) proof of Spencer's original result.

\begin{thm}\label{thm:spencer}
For any set system $(V,\calS)$ with $|V| = n$, $|\calS| = m$, there exists a randomized algorithm running in time $\tilde{O}((n+m)^3)$\,\footnote{Throughout, $\tilde{O}(\,)$ hides polylogarithmic factors.} that with probability at least $1/2$, computes a coloring $\chi:V \rgta \dpm$ such that $\chi(\calS) < K \sqrt{n \cdot \log_2(m/n)}$, where $K$ is a universal constant.
\end{thm}
The constant $K$ above can be taken as $13$ for the case of $m = n$. Observe that our bound matches Spencer's result for all ranges of $m,n$, whereas Bansal's result loses an additional factor of $\Omega(\sqrt{\log(m/n)})$.

We also get a similar constructive proof of Srinivasan's result~\cite{Srinivasan97} for minimizing discrepancy in the ``Beck-Fiala Setting'' where each variable is constrained to occur in a bounded number of sets. Bansal was able to use his SDP based approach to give a constructive proof of Srinivasan's result. Our techniques for Theorem~\ref{thm:spencer} also extend to this setting matching the best known constructive bounds.
\begin{thm}\label{thm:srinivasan}
Let $(V,\calS)$ be a set-system with $|V| = n$, $|\calS| = m$ and each element of $V$ contained in at most $t$ sets from $\calS$. Then, there exists a randomized algorithm running in time $\tilde{O}((n+m)^5)$ that with probability at least $1/2$ computes a coloring $\chi:V \rgta \dpm$ such that $\chi(\calS) < K \sqrt{t} \cdot \log n$, where $K$ is a universal constant.
\end{thm}

We remark that non-constructively, a better bound of $O(\sqrt{t \cdot \log n})$ was obtained by Banaszczsyk~\cite{Banaszczyk98} using techniques from convex geometry. Beck and Fiala~\cite{BeckFiala81} proved that $\disc(\calS)< 2t$ and
conjectured that $\disc(\calS) = O(\sqrt{t})$ and this remains a major open problem in discrepancy minimization.

\section{Outline of Algorithm}
To describe the algorithm we first set up some notation. Fix a set system $(V,\calS)$ with $V = \{1,\ldots,n\}$ and $|\calS| = m$. As is usually done, we shall assume that $m \geq n$ -- the general case can be easily reduced to this situation. Similar to Spencer's original proof our algorithm also works by first finding a ``partial coloring'': $\chi:V \rgta [-1,1]$ such that
\begin{itemize}
\item For all $S \in \calS$, $|\chi(S)| = O(\sqrt{n \log(m/n)})$.
\item $|\{i:|\chi(i)| = 1\}| \geq cn$, for a fixed constant $c > 0$.
\end{itemize}
Given such a partial coloring, we can then recurse (as in Spencer's original proof) by running the algorithm on the set of variables assigned values in $(-1,1)$ without changing the colors of variables assigned values in $\dpm$. Eventually, we will converge to a full coloring and the total discrepancy (a geometrically decreasing series with ratio roughly $\sqrt{1-c}$) can be bounded by $O(\sqrt{n \log(m/n)})$. Henceforth, we will focus on obtaining such a partial coloring.

Let $v_1,\ldots,v_m \in \R^n$ be the indicator vectors of the sets in $\calS$. Then, the discrepancy of $\chi$ on $\calS$ is $\chi(\calS)= \max_{i \in [m]}|\ip{\chi,v_i}|$. Our partial coloring algorithm (as does Spencer's approach) works in the more general context of arbitrary vectors, and we will work in this general context.

\begin{thm}[Main Partial Coloring Lemma]\label{thm:update}
Let $v_1,\ldots,v_m \in \R^n$ be vectors, and $x_0 \in [-1,1]^n$ be a ``starting'' point. Let $c_1,\ldots,c_m \ge 0$ be thresholds such that $\sum_{j=1}^m \exp(-c_j^2/16) \le n/16$. Let $\delta>0$ be a small approximation parameter. Then there exists an efficient randomized algorithm which with probability at least $0.1$ finds a point $x \in [-1,1]^n$ such that
\begin{enumerate}
\item[(i)] $|\ip{x-x_0, v_j}| \le c_j \|v_j\|_2$.
\item[(ii)] $|x_i| \ge 1-\delta$ for at least $n/2$ indices $i \in [n]$.
\end{enumerate}
Moreover, the algorithm runs in time $O((m+n)^3 \cdot \delta^{-2} \cdot \log(nm/\delta))$.
\end{thm}

Note that the probability of success $0.1$ can be boosted by simply running the algorithm multiple times. Given the above result, we can get the desired partial coloring needed for minimizing set discrepancy by applying the theorem to the indicator vectors of the sets $S \in \calS$ with $\delta = 1/n$, and $x_0 = \mathsf{0}^n$. Combining the above with the recursive analysis gives Theorem~\ref{thm:spencer} with a running time of $\tilde{O}((n+m)^5)$. It was pointed to us by Spencer that we can in fact take $\delta = 1/\log n$ and then use randomized rounding to get the running time stated in Theorem~\ref{thm:spencer}.

We stress that Spencer's original approach shows the existence of a true partial coloring (the colors take values in $\{-1,0,1\}$), whereas our approach gives a fractional coloring---the colors take values in $[-1,1]$ though many of the colors are close to $\{-1,1\}$. 

The constructive proof of Srinivasan's result, Theorem~\ref{thm:srinivasan}, follows a similar outline starting from our partial coloring lemma. We defer the details to Section \ref{sec:discmin}.

We now describe the proof of the partial coloring lemma.

\subsection{Partial Coloring by Walking on The Edge}\label{sec:walkintro}
We will find the desired vector $x$ by performing a constrained random walk that we refer to as {\it Edge-Walk} for reasons that will become clear later.

We first describe the algorithm conceptually, ignoring the approximation parameter $\delta$. We will assume throughout that $\|v_1\|_2=\ldots=\|v_m\|_2=1$ as this normalization does not change the problem. Consider the following polytope $\PP$ which describes the legal values for $x \in \R^n$,
$$
\PP:=\{x \in \R^n: |x_i| \le 1 \; \forall i \in [n],\; |\ip{x-x_0,v_j}| \le c_j \; \forall j \in [m]\}.
$$
We will refer to the constraints $|x_i| \le 1$ as {\em variable constraints} and to the constraints $|\ip{x-x_0,v_j}| \le c_j$ as {\em discrepancy constraints}. The partial coloring lemma can be rephrased in terms of the polytope $\PP$ as follows: there exists a point $x \in \PP$ that satisfies at least $n/2$ variable constraints without any slack. Intuitively, this corresponds to finding a point $x$ in $\PP$ that is as far away from origin as possible; the hope being that if $\|x\|_2$ is large, then in fact many of the coordinates of $x$ will be close to $1$ in absolute value.
We find such a point (and show it's existence) by simulating a constrained Brownian motion in $\PP$. (If uncomfortable with Brownian motion, the reader can view the walk as taking very small discrete Gaussian steps, which is what we will do in the actual analysis.)

Consider a random walk in $\PP$ corresponding to the Browninan motion starting at $x=x_0$. Whenever the random walk reaches a face of the polytope, it continues inside this face. We continue the walk until we reach a vertex $x \in \PP$. The idea being that we want to get away from origin, but do not want to cross the polytope -- so whenever a constraint (variable or discrepancy) becomes tight we do not want to change the constraint and continue in the subspace orthogonal to the defining constraint. We call this random walk the ``Edge-Walk'' in $\PP$.

By definition, the random walk is constrained to $\PP$, and $|\ip{x-x_0,v_j}| \le c_j$ for all $j \in [m]$. We show that as long as $\sum \exp(-c_j^2) \ll n$, the random walk hits many variable constraints with good probability. That is, the end vertex $x$ has $x_i \in \{-1,1\}$ for many indices. This step relies on a martingale tail bound for Gaussian variables and an implicit use of the $\ell_2$-norm as a potential function for gauging the number of coordinates close to $1$ in absolute value.

The actual algorithm differs slightly from the above description. First, we will not run the walk until we reach a vertex of $\PP$, but after a certain `time' has passed, which will still guarantee the above conditions. Second, we will approximate the continuous random walk by many small discrete steps.

\section{Comparison with Entropy Method}
Here we contrast our result with Beck's partial coloring lemma~\cite{Beck81} based on the Entropy method which has many applications in discrepancy theory. While similar in spirit, our partial coloring lemma is incomparable and in particular, even the existence of the vector $x$ as in Theorem~\ref{thm:update} does not follow from Beck's partial coloring lemma. 
 
We first state Beck's partial coloring lemma as formulated in \cite{Matousek98}.
\begin{thm}[Entropy Method]\label{thm:beckpcl}
Let $(V,\calS)$ be a set-system with $V = \{1,\ldots,n\}$. Let $\Delta:\calS \rightarrow \R_+$ be such that $\sum_{S \in \calS} g(\Delta_S/\sqrt{|S|}) \leq n/5$, where $g:\R_+ \rightarrow \R_+$ is defined by,
\[ g(\lambda) =
\begin{cases}
  K e^{-\lambda^2/9}, & \lambda > 0.1\\
  K \ln(1/\lambda), & \lambda \leq 0.1
\end{cases},\]
where $K$ is an absolute constant. Then, there exists $\chi \in \{-1,0,1\}^n$ with $|\{i: \chi_i \neq 0\}| \geq n/2$ such that $|\sum_{i \in S} \chi_i| \leq \Delta_S$ for every $S \in \calS$.
\end{thm}
\newcommand{\poly}{\mathrm{poly}}
By applying our Theorem~\ref{thm:update} to the indicator vectors of the sets in $\calS$ and $\delta = 1/\poly(n)$ sufficiently small we get the following corollary.
\begin{cor}\label{thm:pcoloring}
  Let $(V,\calS)$ be a set-system with $V = \{1,\ldots,n\}$. Let $\Delta:\calS \rightarrow \R_+$ be such that
$$\sum_{S \in \calS} \exp(-\Delta_S^2/16|S|) \leq n/16.$$
Then, there exists $\chi \in [-1,1]^n$ with $|\{i: |\chi_i| = 1\}| \geq n/2$, such that $|\sum_{i \in S} \chi_i| \leq \Delta_S + 1/\poly(n)$, for every $S \in \calS$. Moreover, there exists a randomized $\poly(|\calS|, n)$-time algorithm to find $\chi$.
\end{cor}
\ignore{
\begin{proof}
Apply Theorem \ref{thm:update} to the indicator vectors of the sets $S \in \calS$ with $\delta = 1/n$, $x_0 = \mathsf{0}^n$ to obtain a vector $x$ as in the statement. Let $\chi \in \R^n$ be the vector obtained by rounding each $x_i$ with $|x_i| \geq 1- \delta$ to $sign(x_i)$. The statement now follows immediately from the conditions of Theorem \ref{thm:update}.
\end{proof}}

The above result strengthens the Entropy method in two important aspects. Firstly, our method is constructive. In contrast, the entropy method is non-constructive and the constructive discrepancy minimization algorithms of Bansal do not yield the full partial coloring lemma as in Theorem~\ref{thm:beckpcl}. Secondly, the above result can tolerate many more {\it stringent constraints} than the Entropy method. For instance, the entropy method can only allow $O(n/\log n)$ of the sets in $\calS$ to have discrepancy $1/n$, whereas our result can allow $\Omega(n)$ of the sets to have such small discrepancy. We believe that this added flexibility in achieving much smaller discrepancy for a constant fraction of sets could be useful elsewhere.

One weakness of Theorem \ref{thm:update} is that we do not strictly speaking get a proper partial coloring: the non $\{1,-1\}$ variables in our coloring $\chi$ can take any value in $(-1,1)$. This however does not appear to be a significant drawback, as Corollary \ref{thm:pcoloring} can also be made to work from an arbitrary {\it starting point} $x_0$ as in the statement of Theorem \ref{thm:update}.


\section{Preliminaries}
We start with some notation and few elementary properties of the Gaussian distributions.

\subsection{Notation} Let $[n]=\{1,\ldots,n\}$. Let $e_1,\ldots,e_n$ denote the standard basis for $\R^n$. We denote
random variables by capital letters and distributions by calligraphic letters. We write $X \sim \mathcal{D}$ for a random variable
$X$ distributed according to a distribution $\mathcal{D}$.

\subsection{Gaussian distribution}
Let $\N(\mu,\sigma^2)$ denote the Gaussian distribution with mean $\mu$ and variance $\sigma^2$.
A Gaussian distribution is called {\em standard} if $\mu=0$ and $\sigma^2=1$. If $G_1 \sim \N(\mu_1,\sigma_1^2)$ and $G_2 \sim \N(\mu_2,\sigma_2^2)$ then for $t_1,t_2 \in \R$ we have
$$
t_1 G_1 + t_2 G_2 \sim \N(t_1 \mu_1+t_2 \mu_2, t_1^2 \sigma_1^2 + t_2^2 \sigma_2^2).
$$

Let $V \subseteq \R^n$ be a linear subspace. We denote by $G \sim \N(V)$ the standard multi-dimensional Gaussian distribution supported on $V$: $G=G_1 v_1 + \ldots + G_d v_d$, where $\{v_1,\ldots,v_d\}$ is an orthonormal basis for $V$ and $G_1,\ldots,G_d \sim \N(0,1)$ are independent standard Gaussian variables. It is easy to check that this definition is invariant of the choice of the basis $\{v_1,\ldots,v_d\}$. We will need the following simple claims.

\begin{claim}\label{claim:projection_unit_vector}
Let $V \subseteq \R^n$ be a linear subspace and let $G \sim \N(V)$. Then, for all $u \in \R^n$, $\ip{G,u} \sim \N(0,\sigma^2)$, where $\sigma^2 \le \|u\|_2^2$.
\end{claim}

\begin{proof}
Let $G=G_1 v_1 + \ldots + G_d v_d$ where $\{v_1,\ldots,v_d\}$ is an orthonormal basis for $V$ and $G_1,\ldots,G_d \sim \N(0,1)$ are independent. Then $\ip{G,u}=\sum_{i=1}^d \ip{u,v_i} \cdot G_i$ is Gaussian with mean zero and variance $\sum_{i=1}^d \ip{u,v_i}^2 \le \|u\|_2^2$.
\end{proof}

\begin{claim}\label{claim:projection_unit_basis}
Let $V \subseteq \R^n$ be a linear subspace and let $G \sim \N(V)$. Let $\ip{G,e_i}\sim \N(0,\sigma_i^2)$. Then $\sum_{i=1}^n \sigma_i^2=\dim(V)$.
\end{claim}

\begin{proof}
Let $G=G_1 v_1+\ldots+G_d v_d$ where $v_1,\ldots,v_d$ are an orthonormal basis for $V$ and $G_1,\ldots,G_d \sim \N(0,1)$ are independent. Then, $\sum_{i=1}^n \sigma_i^2 = \sum_{i=1}^n \E[|\ip{G,e_i}|^2] = \E[\|G\|_2^2] = \sum_{i=1}^d \|v_i\|_2^2 \cdot \E[G_i^2]= d=\dim(V).
$
\end{proof}

The following is a standard tail bound for Gaussian variables.
\begin{claim}\label{claim:gaussian_tail}
Let $G \sim N(0,1)$. Then, for any $\lambda>0$, $\Pr[|G| \ge \lambda] \le 2 \exp(-\lambda^2/2)$.
\end{claim}
\ignore{
We will also need the following anti-concentration bound.\onote{apparently never used... remove?}
\begin{claim}\label{claim:guassian_anti_concentration}
Let $G \sim N(0,1)$. The for any $t<1$ we have $\Pr[|G| \le t] \le 2t$.
\end{claim}}
We will also need the following tail bound on martingales with Gaussian steps. It is a mild generalization of Lemma 2.2 in \cite{Bansal10} and we omit the proof.

\begin{lemma}[\cite{Bansal10}]\label{lemma:martingale}
Let $X_1,\ldots,X_T$ be random variables. Let $Y_1,\ldots,Y_T$ be random variables where each $Y_i$ is a function of $X_i$. Suppose that for all $1 \le i \le T$, $x_1,\ldots,x_{i-1} \in \R$, $Y_i | (X_1 = x_1, X_2 = x_2,\ldots,X_{i-1} = x_{i-1})$ is Gaussian with mean zero and variance at most one (possibly different for each setting of $x_1,\ldots,x_{i-1}$). Then for any $\lambda>0$,
$$
\Pr[|Y_1+\ldots+Y_T| \ge \lambda \sqrt{T}] \le 2 \exp(-\lambda^2/2).
$$
\end{lemma}
\ignore{
\begin{lemma}\label{lemma:martinagle}
Let $X_1,\ldots,X_T$ be random variables. Let $Y_1,\ldots,Y_T$ be random variables where $Y_i$ is a function of $X_i$, so that for all $1 \le i \le T$ we have that $Y_i|(X_1,\ldots,X_{i-1})$ is Gaussian with mean zero and variance $\sigma_i^2 \le 1$. Note that $\sigma_i$ may depend on $X_1,\ldots,X_{i-1}$. Then for any $\lambda>0$,
$$
\Pr[|Y_1+\ldots+Y_N| \ge \lambda \sqrt{N}] \le 2 \exp(-\lambda^2/2).
$$
\end{lemma}}

\section{Main Partial Coloring Lemma}
We are now ready to present our main {\it partial coloring} algorithm and prove Theorem~\ref{thm:update}. We shall use the notation from the theorem statement and Section~\ref{sec:walkintro}.

Let $\gamma>0$ be a small step size so that $\delta=O(\gamma \sqrt{\log(nm/\gamma)})$. We note that the correctness of the algorithm is not affected by the choice of $\gamma$, as long as it is small enough; only the running time is affected.

Let $T = K_1/\gamma^2$, where $K_1 = 16/3$. We assume that $\delta < 0.1$. The algorithm will produce intermediate steps $X_0=x_0,X_1,\ldots,X_T \in \R^n$ according to the following update process\footnote{We call the random walk ``Edge-Walk'' because geometrically, once the walk (almost) hits an edge (face) of the polytope $\PP$, it stays on the edge.}

\paragraph{Edge-Walk:}For $t = 1,\ldots,T$ do
\begin{itemize}
\item Let $\Cvar_t := \Cvar_t(X_{t-1}) = \{i \in [n]: |(X_{t-1})_i| \ge 1-\delta\}$ be the set of variable constraints `nearly hit' so far.
\item Let $\Crow_t := \Crow_t(X_{t-1}) = \{j \in [m]: |\ip{X_{t-1}-x_0,v_j}| \ge c_j-\delta\}$ be the set of discrepancy constraints `nearly hit' so far.
\item Let
$ \O_t := \O(X_{t-1}) = \{u \in \R^n: u_i=0\; \forall i \in \Cvar_t,\quad \ip{u,v_j}=0 \;\forall j \in \Crow_t\}$ be the linear subspace orthogonal to the `nearly hit' variable and discrepancy constraints.
\item Set $X_t := X_{t-1} + \gamma U_t$, where $U_t \sim \N(\O_t)$.
\end{itemize}

\ignore{
Let $T=100/\gamma^2$ denote the total number of steps in the discrete random walk. The algorithm will produce intermediate steps $X_0=x_0,X_1,\ldots,X_T \in \R^n$ according to the following update steps. We define how $X_t$ is generated given $X_{t-1}$. Define the set of variable constraints `nearly hit' by the random walk so far
$$
\Cvar_t := \Cvar_t(X_{t-1}) = \{i \in [n]: |(X_{t-1})_i| \ge 1-\delta\}
$$
and the set of discrepancy constraints `nearly hit' by the random walk before step $t$
$$
\Crow_t := \Crow_t(X_{t-1}) = \{j \in [m]: |\ip{X_{t-1}-x_0,v_j}| \ge c_j-\delta\}.
$$
Note that if $|(x_0)_i| \ge 1 - \delta$, then $i \in \Cvar_t$ for all $t \le T$. Similarly, if $c_j \le \delta$ then $j \in \Crow_t$ for all $t \le T$. Given $X_{t-1}$, the next step of the random walk will be in the vector space orthogonal to all the constrains `nearly hit' before step $t$
$$
\O_t = \O(X_{t-1}) := \{u \in \R^n: u_i=0\; \forall i \in \Cvar_t,\quad \ip{u,v_j}=0 \;\forall j \in \Crow_t\}.
$$
We define the update step $U_t \sim \N(\O_t)$ to be a Gaussian variable distributed in $\O_t$ and set
$$X_{t}:=X_{t-1}+ \gamma U_t.$$}
The following lemma captures the essential properties of the random walk.

\begin{lemma}\label{lemma:update_analysis}
Consider the random walk described above. Assume that $\sum_{j=1}^m \exp(-c_j^2/16) \le n/16$. Then, with probability at least $0.1$,
\begin{enumerate}
\item $X_0,\ldots,X_T \in \PP$.
\item $|(X_T)_i| \ge 1-\delta$ for at least $n/2$ indices $i \in [n]$.
\end{enumerate}
\end{lemma}
Theorem~\ref{thm:update} follows immediately from Lemma~\ref{lemma:update_analysis} by setting $x=X_T$. Note that computing $\Cvar_t, \Crow_t$, given $X_{t-1}$ takes time $O(nm)$. Further, once we know the set of constraints defining $\O_t$, we can sample from $\N(\O_t)$ in time $O((n+m)^3)$ by first constructing an orthogonal basis $U$ for $\O_t$ and setting $U_t = \sum_{u \in U} G_u u$, where $G_u \sim \N$ are chosen independently.

We prove Lemma~\ref{lemma:update_analysis} in the remainder of this section. We start with a simple observation that $\Cvar_t,\Crow_t$ can only increase during the random walk.

\begin{claim}\label{claim:monotonicity_of_constraints}
For all $t<T$ we have $\Cvar_t \subseteq \Cvar_{t+1}$ and $\Crow_{t} \subseteq \Crow_{t+1}$. In particular, for $1 \le t < T$, $dim(\O_t) \geq dim(\O_{t+1})$.
\end{claim}

\begin{proof}
Let $i \in \Cvar_t$. That is, $|(X_{t-1})_i| \ge 1-\delta$. Then by definition of the random walk, $U_t \in \O_t$ and $(U_t)_i=0$. Thus, $(X_{t})_i=(X_{t-1})_i$ and $i \in \Cvar_{t+1}$. The argument for discrepancy constraints is analogous.
\end{proof}

We next show that the walk stays inside $\PP$ with high probability.

\begin{claim}\label{claim:points_in_polytope}
For $\gamma \le \delta/\sqrt{C \log(mn/\gamma)}$ and $C$ a sufficiently large constant, with probability at least $1 - 1/(mn)^{C-2}$, $X_0,\ldots,X_T \in \PP$.
\end{claim}

\begin{proof}
The proof involves a simple application of the tail bound from Claim~\ref{claim:gaussian_tail}. Clearly $X_0=x_0 \in \PP$. Let $E_t:=\{X_{t} \notin \PP|X_0,\ldots,X_{t-1} \in \PP\}$ denote the event that $X_t$ is the first element outside $\PP$, so $\Pr[X_0,\ldots,X_T \in \PP] = 1-\sum_{t=1}^{T} \Pr[E_t]$. 

In order to calculate $\Pr[E_t]$, note that if $E_t$ holds then $X_{t}$ must violate either a variable constraint or a discrepancy constraint. Assume for example that $X_t$ violates a variable constraint, say $(X_t)_i > 1$. Since $X_{t-1} \in \PP$ we must have $(X_{t-1})_i \le 1$. However, we we must in fact have $|(X_{t-1})_i| \le 1-\delta$ as otherwise we would have $i \in \Cvar_t$ and hence $(U_t)_i=0$ and $(X_t)_i = (X_{t-1})_i$. Thus, in order for this situation to occur we must have that $|(U_t)_i| \ge \delta/\gamma$. We will show this is very unlikely.

Let $W:=\{e_1,\ldots,e_n,v_1,\ldots,v_m\}$. We conclude that if $E_t$ holds
then $|\ip{X_t-X_{t-1},w}| \ge \delta$ for some $w \in W$. That is, $|\ip{U_t,w}| \ge \delta/\gamma$. We next bound the probability of these events.
Since $U_t \sim \N(\O_t)$ we have by Claim~\ref{claim:projection_unit_vector} that $\ip{U_t,w}$ is Gaussian with mean $0$ and variance at most $1$. Hence by Claim~\ref{claim:gaussian_tail},
$$
\Pr[|\ip{U_{t},w}| \ge \delta/\gamma] \le 2 \exp(-(\delta/\gamma)^2/2).
$$
By our setting of parameters $\delta/\gamma = \sqrt{C \log(nm/\gamma)})$ and $T=O(1/\gamma^2)$. Thus,
\begin{align*}
\Pr[X_0,\ldots,X_T \notin \PP] &= \sum_{t=1}^T \Pr[E_t] \le \sum_{t=1}^T \sum_{w \in W} \Pr[|\ip{U_t,w}| \ge\delta/\gamma] \le T \cdot (nm) \cdot \frac{\gamma^2}{(mn)^C} \le \frac{1}{(mn)^{C-2}},
\end{align*}
for $C$ large enough.
\end{proof}

We are now ready to prove Lemma \ref{lemma:update_analysis}. The intuition behind the proof is as follows. We first use the hypothesis on the thresholds $c_j, j \in [m]$, to argue that $\E[\,|\Crow_T|\,] \ll  n$. This follows from the definition of the walk and a simple application of the martingale tail bound of Lemma~\ref{lemma:martingale}. Note that to prove the lemma it essentially suffices to argue that $\E[|\Cvar_T|] = \Omega(n)$ (we can then use Markov's inequality). Roughly speaking, we do so by a ``win-win'' analysis. Consider an intermediate update step $t \le T$. Then, either $|\Cvar_t|$ is large, in which case we are done, or $|\Cvar_t|$ is small in which case $dim(\O_{t-1})$ is large so that $\E[\|X_t\|^2]$ increases significantly (with noticeable probability) due to Claim~\ref{claim:projection_unit_basis}. On the other hand, $\|X_t\|^2 \le n$ as all steps stay within the polytope $\PP$ (with high probability). Hence, $|\Cvar_t|$ cannot be small for many steps and in particular $|\Cvar_T|$ will be large with noticeable probability.

We first argue that $\E[\,|\Crow_T|\,]$ is small. That is, on average only a few discrepancy constraints are ever nearly hit.


\begin{claim}\label{claim:few_row_constraints}
$\E[|\Crow_T|] < n/4$. 
\end{claim}
\begin{proof}
Let $J:=\{j: c_j \le 10 \delta\}$. To bound the size of $J$, we have
$$
n/16 \ge \sum_{j \in J} \exp(-c_j^2/16) \ge |J| \cdot \exp(-100 \delta^2/16) \geq |J| \cdot \exp(-1/16) > 9 |J|/10,
$$
and hence $|J| \le 1.2 n/16$. Now, for $j \notin J$, if $j \in \Crow_T$, then $|\ip{X_T-x_0,v_j}| \ge c_j-\delta \ge 0.9 c_j $. We will bound the probability that this occurs. Recall that $X_T=x_0+\gamma(U_1+\ldots+U_T)$ and define $Y_i=\ip{U_i,v_j}$. Then, for $j \notin J$, we have
$$
\Pr[j \in \Crow_T] \le \Pr[\,|Y_1+\ldots+Y_T| \ge 0.9 c_j/\gamma\,].
$$
We next apply Lemma~\ref{lemma:martingale}. Note that the conditions of the lemma apply, since $U_1,\ldots,U_T$ is a sequence of random variables, $Y_i$ is a function of $U_i$ and $Y_i|(U_1,\ldots,U_{i-1})$ is Gaussian with mean zero and variance at most one (by Claim~\ref{claim:projection_unit_vector}). Hence,
$$
\Pr[j \in \Crow_T] \le 2\exp(-(0.9 c_j)^2/ 2 \gamma^2 T) = 2\exp(-(0.9 c_j)^2/2 K_1 T) < 2\exp(-c_j^2/ 16).
$$
So
$$
\E[|\Crow_T|] \le |J|+\sum_{j \notin J}\Pr[j \in \Crow_T] \le 1.2 n/16 + 2n/16 < n/4.$$
\end{proof}


\begin{claim}\label{claim:norm_xT}
$\E[\|X_T\|_2^2] \le n$.
\end{claim}

\begin{proof}
We will show that $\E[(X_T)_i^2] \le 1$ for all $i \in [n]$. Conditioning on the first $t$ for which $i \in \Cvar_t$ (or that no such $t$ exists), we get
$$
\E[(X_T)_i^2] = \Pr[i \notin \Cvar_T]\, \E[(X_T)_i^2|i \notin \Cvar_T]+\sum_{t=1}^T  \Pr[i \in \Cvar_t \setminus \Cvar_{t-1}]\,\E[(X_T)_i^2|i \in \Cvar_t \setminus \Cvar_{t-1}].
$$
Clearly $\E[(X_T)_i^2|i \notin \Cvar_T] \le 1$. For $t \le T$, we have
$$
\E[(X_T)_i^2|i \in \Cvar_t \setminus \Cvar_{t-1}] = \E[(X_t)_i^2|i \in \Cvar_t \setminus \Cvar_{t-1}] \le 1-\delta+\gamma \E[|(U_t)_i|_2^2] \le 1,
$$
where we used the fact that $(U_t)_i$ is a Gaussian variable with mean zero and variance at most one (by Claim~\ref{claim:projection_unit_vector}).
\end{proof}

Finally, we show that $\E[|\Cvar_T|]$ is large. That is, on average we will nearly hit a constant fraction of the variable constraints.
\begin{claim}\label{claim:many_constrains}
$\E[\,|\Cvar_T|\,] \ge 0.56 n$. 
\end{claim}
\begin{proof}
We start by computing the average norm of $X_t$.
$$
\E[\|X_t\|_2^2] = \E[\|X_{t-1}+\gamma U_t\|_2^2] = \E[\|X_{t-1}\|_2^2]+\gamma^2 \E[\|U_t\|_2^2]=\E[\|X_{t-1}\|_2^2] + \gamma^2 \E[\dim(\O_t)],
$$
where we used that fact that given $X_{t-1}$, $\E[U_t|X_{t-1}]=0$ and $\E[\|U_t\|_2^2|X_{t-1}]=\dim(\O_t)$, by Claim~\ref{claim:projection_unit_basis}. Hence, by Claim \ref{claim:norm_xT},
$$
n \ge \E[\|X_T\|_2^2] \ge \gamma^2 \sum_{t=1}^T \E[\dim(\O_t)] \ge \gamma^2 |T| \cdot \E[\dim(\O_T)] = K_1 \cdot \E[\dim(\O_T)] = K_1 \E[(n - |\Cvar_T| - |\Crow_T|)].
$$
Therefore, $\E[|\Cvar_T|] \geq n(1-1/K_1) - \E[|\Crow_T|] \geq n (1 - 1/K_1 - 1/4) > (0.56)n$, where the second inequality follows from Claim~\ref{claim:few_row_constraints}.
\end{proof}

\ignore{
\begin{claim}\label{claim:many_constrains}
With probability at least $0.9$, we have that $|\Cvar_T|+|\Crow_T| \ge 0.9n$.
\end{claim}

\begin{proof}
We will prove the equivalent statement $\Pr[\dim(\O_T) \ge 0.1n] \le 0.1$. We start by computing the average norm of $X_t$.
$$
\E[\|X_t\|_2^2] = \E[\|X_{t-1}+\gamma U_t\|_2^2] = \E[\|X_{t-1}\|_2^2]+\gamma^2 \E[\|U_t\|_2^2]=\E[\|X_{t-1}\|_2^2] + \gamma^2 \E[\dim(\O_t)],
$$
where we used that fact that given $X_{t-1}$ we have that $\E[U_t|X_{t-1}]=0$ and $\E[\|U_t\|_2^2|X_{t-1}]=\dim(\O_t)|X_{t-1}$.
By Claim~\ref{claim:norm_xT} we know that $\E[\|X_T\|_2^2|] \le n$. Hence
$$
n \ge \E[\|X_T\|_2^2] \ge \gamma^2 \sum_{t=1}^T \E[\dim(O_t)] \ge \gamma^2 |T| \cdot \E[\dim(\O_T)] = 100 \cdot \E[\dim(\O_T)].
$$
We thus have by the Markov inequality that
$$
\Pr\left[\dim(\O_T) \ge 0.1n\right] \le 0.1.
$$
\end{proof}
}
Lemma~\ref{lemma:update_analysis} now follows immediately from Claim \ref{claim:points_in_polytope} and Claim \ref{claim:many_constrains}.

\begin{proof}[Proof of Lemma~\ref{lemma:update_analysis}]
From Claim \ref{claim:many_constrains} and the fact that $|\Cvar_T| \le n$, it follows that $\P[|\Cvar_T| \geq n/2] \geq 0.12$. Combining with Claim \ref{claim:points_in_polytope}, with probability at least $0.12 - 1/poly(m,n) > 0.1$, $|\Cvar_T| \geq n/2$ and $X_T \in \PP$ which shows the lemma.
\ignore{
By claims \ref{claim:points_in_polytope}, \ref{claim:few_row_constraints} and \ref{claim:many_constrains} we know that with probability at least $0.7$ we have that
\begin{itemize}
\item $X_0,\ldots,X_T \in \PP$.
\item $|\Crow_T| \le 0.1n$.
\item $|\Cvar_T|+|\Crow_T| \ge 0.9n$.
\end{itemize}
Hence $|\Cvar_T| \ge 0.8n$, that is for at least $0.8n$ indices $i \in [n]$ we have $|(X_T)_i| \ge 1-\delta$.}
\end{proof}

\section{Discrepancy Minimization from Partial Coloring}\label{sec:discmin}
We now derive Theorem \ref{thm:spencer} and Theorem \ref{thm:srinivasan} from our partial coloring lemma.
\begin{proof}[Proof of Theorem \ref{thm:spencer}]
Let $(V,\calS)$ be a system with $|V| = n$ and $|\calS|=m$. Let $v_1,\ldots,v_m \in \R^n$ be the indicator vectors of the sets in $\calS$. We set $\delta = 1/(8 \log m)$. Let $\alpha(m,n) = 8\sqrt{\log(m/n)}$. Then, $m \cdot \exp(-\alpha(m,n)^2/16) < n/16$. Therefore, by Theorem \ref{thm:update} applied to $v_1,\ldots,v_m$ and starting point $x_0 = \mathsf{0}^n$, with probability at least $0.1$ we find a vector $x_1 \in [-1,1]^n$ such that $|\ip{v_j,x_1}| < \sqrt{n} \cdot \alpha(m,n)$ for all $j \in m$ and $|\{i: |(x_1)_i| \geq 1- \delta\}| \geq n/2$. We can boost this probability further by repeating the algorithm $O(\log n)$ times; from now on we will ignore the probability that the algorithm does not find such a vector.

Let $I_1 = \{i: |(x_1)_i| < 1 - \delta\}$ be the coordinates not `fixed' in the first step and set $n_1 = |I_1|$. We now iteratively apply Theorem \ref{thm:update} to the restricted system described by the vectors $v_1^1 = (v_1)_{I_1},\ldots, v_m^1 = (v_m)_{I_1} \in \R^{n_1}$ and starting point $(x_1)_{I_1}$ to get another vector $x_2 \in [-1,1]^{n_1}$ such that $|\langle v_j^1,x_2\rangle| < \sqrt{n_1} \cdot \alpha(m,n_1)$ for all $j \in [m]$ and $|\{i: |(x_2)_i| \geq 1- \delta\}| \geq n_1/2$. By iterating this procedure for at most $t = 2 \log n$ times and concatenating the resulting vectors appropriately we get $x \in \R^n$ such that $|x_i| \geq 1 - \delta$ for all $i \in [n]$ and for every $j \in [m]$,
\begin{align*}
 |\ip{v_j, x}| &< \sqrt{n} \cdot \alpha(m,n) + \sqrt{n_1} \cdot \alpha(m,n_1) + \cdots + \sqrt{n_t} \cdot \alpha(m,n_t)\\
&< \sqrt{n} \sum_{r = 0}^\infty  \frac{ 8 \sqrt{\log(m \cdot 2^r/n)}}{2^{r/2}}\\
&< C\sqrt{n \cdot \log(m/n)},
\end{align*}
for $C$ a universal constant.

We now round $x$ to get a proper coloring $\chi \in \dpm^n$. Let $\chi \in \dpm^n$ be obtained from $x$ as follows: for $i \in [n]$, $\chi_i = sign(x_i)$ with probability $(1+|x_i|)/2$ and $-sign(x_i)$ with probability $(1 - |x_i|)/2$, so that $\E[\chi_i]=x_i$. Let $Y=\chi-x$.
Fix some $j \in [m]$. Then, the discrepancy of $\chi$ with $v_j$ is  
$$
|\ip{\chi,v_j}| \le |\ip{x,v_j}|+|\ip{Y,v_j}| \le C \sqrt{n \log(m/n)} + |\ip{Y,v_j}|.
$$
We will show that with high probability, $|\ip{Y,v_j}| \le \sqrt{n}$ for all $1 \le j \le m$. Fix some $j \in [m]$ and consider $\ip{Y,v_j}$. We have that $|Y_i| \le 2$, $\E[Y_i]=0$ and $\Var(Y_i) \le \delta$. We also have $\|v_j\|_2 \le \sqrt{n}$ and $\|v_j\|_{\infty} \le 1$. Thus, by a standard Chernoff bound (see e.g., Theorem 2.3 in \cite{ChungLu}),
$$
\P\left[ |\ip{Y,v_j}| > 2 \sqrt{2\log m} \cdot \sqrt{n \delta} \right] \leq 2 \exp(-2 \log m) < 1/2m.
$$
Therefore, by the union bound and our choice of $\delta$, with probability at least $1/2$ we have that $|\ip{Y,v_j}| \le \sqrt{n}$ for all $1 \le j \le m$. Therefore, $|\ip{\chi,v_j}| \le C\sqrt{n \log(m/n)}+\sqrt{n}$ for all $1 \le j \le m$.

The running time is dominated by the $O(\log^2 n)$ uses of Theorem~\ref{thm:update}. Thus, the total running time is $O((n+m)^3 \log^5(mn)) = \tilde{O}((n+m)^3)$. 
\end{proof}
The constant in the theorem can be sharpened to be $13$ by fine tuning the parameters. We do not dwell on this here. We next prove Theorem \ref{thm:srinivasan}.

\begin{proof}[Proof of Theorem \ref{thm:srinivasan}]
The proof is similar to the above argument and we only sketch the full proof. Set $\delta = 1/n$. Let $(V,\calS)$ be the set system. Let $v_1,\ldots,v_m$ be the indicator vectors of the sets in $\calS$ and let $c_j = C \sqrt{t}/\|v_j\|_2$ for $C$ to be chosen later. Observe that $\sum_j \|v_j\|_2^2 \leq n t$ as each element appears in at most $t$ sets. In particular, the number of vectors $v_j$ with $\|v_j\|_2^2$ in $[2^r t, 2^{r+1}t]$ is at most $n/2^r$. Therefore,
\[ \sum_j \exp(-c_j^2/16) < \sum_{r=0}^\infty \frac{n \cdot \exp(-C^2/16 \cdot 2^{r+1})}{2^r} < n/16,\]
for $C$ a sufficiently large constant. Thus, by applying Theorem \ref{thm:update} to the vectors $v_j$ and thresholds $c_j$ for $j \in [m]$, with probability at least $0.1$ we get a vector $x_1 \in [-1,1]^n$ such that $|\ip{v_j, x_1}| < C \sqrt{t}$ for all $j \in [m]$ and $|\{i: |(x_1)_i| \geq 1 - \delta\}| > n/2$.

By iteratively applying the same argument as in the proof of Theorem \ref{thm:update} for $2 \log n$ steps, we get a vector $x \in [-1,1]^n$ with $|x_i| \geq 1- \delta$ for all $i$ and $|\ip{v_j, x}| < 2 C \sqrt{t} \log n$ for all $j \in [m]$. The theorem now follows by rounding the $x$ to the nearest integer coloring $\chi$: $\chi_i = sign(x_i)$ for all $i \in [m]$.
\end{proof}
\paragraph{Acknowledgments} We would like to thank Oded Regev for many discussions and collaboration at the early stages of this work. We thank Joel Spencer for his encouragement and enthusiasm about this work: part of our presentation is inspired by a lecture he gave on this result at the Institute for Advanced Study, Princeton. We also thank him for the observation on improving the run time of Theorem \ref{thm:update} and allowing us to include it here. We thank Nikhil Bansal for valuable comments and discussions.

\bibliographystyle{alpha}
\bibliography{disc_min_alg}

\begin{thebibliography}{Mat99}

\bibitem[AS11]{AlonS11}
N.~Alon and J.H. Spencer.
\newblock {\em The Probabilistic Method}.
\newblock Wiley Series in Discrete Mathematics and Optimization. John Wiley \&
  Sons, 2011.

\bibitem[Ban98]{Banaszczyk98}
Wojciech Banaszczyk.
\newblock Balancing vectors and gaussian measures of n-dimensional convex
  bodies.
\newblock {\em Random Struct. Algorithms}, 12(4):351--360, 1998.

\bibitem[Ban10]{Bansal10}
Nikhil Bansal.
\newblock Constructive algorithms for discrepancy minimization.
\newblock In {\em FOCS}, pages 3--10, 2010.

\bibitem[Bec81]{Beck81}
J.~Beck.
\newblock {R}oths estimate of the discrepancy of integer sequences is nearly
  sharp.
\newblock {\em Combinatorica}, 1(4):319--325, 1981.

\bibitem[BF81]{BeckFiala81}
J.~Beck and T.~Fiala.
\newblock Integer-making theorems.
\newblock {\em Discrete Applied Mathematics}, 3(1):1--8, 1981.

\bibitem[Cha02]{Chazelle2002}
B.~Chazelle.
\newblock {\em The Discrepancy Method: Randomness and Complexity}.
\newblock Cambridge University Press, 2002.

\bibitem[CL06]{ChungLu}
Fan Chung and Linyuan Lu.
\newblock {\em {Complex Graphs and Networks}}.
\newblock American Mathematical Society, 2006.

\bibitem[Mat98]{Matousek98}
Jir\'{\i} Matou\v{s}ek.
\newblock An l$_{\mbox{p}}$ version of the beck-fiala conjecture.
\newblock {\em Eur. J. Comb.}, 19(2):175--182, 1998.

\bibitem[Mat99]{Matousek1999}
J.~Matou{\v{s}}ek.
\newblock {\em Geometric Discrepancy: An Illustrated Guide}.
\newblock Algorithms and Combinatorics. Springer, 1999.

\bibitem[Spe85]{Spencer85}
Joel Spencer.
\newblock Six standard deviations suffice.
\newblock {\em Transactions of the American Mathematical Society},
  289(2):679--706, 1985.

\bibitem[Sri97]{Srinivasan97}
A.~Srinivasan.
\newblock Improving the discrepancy bound for sparse matrices: Better
  approximations for sparse lattice approximation problems.
\newblock In {\em ACM-SIAM Symposium on Discrete Algorithms}, pages 692--701,
  1997.

\end{thebibliography}

\end{document}